\documentclass{basi}
\usepackage[british]{babel}
\usepackage[varg]{txfonts}
\usepackage{rotating}
\usepackage{dcolumn}
\begin{document}
\title{Transient induced MHD oscillations : A tool to probe the solar active regions\thanks{By
            A.K. Srivastava, email: \texttt{aks@aries.res.in}}}
\author[A.~K.~Srivastava~et~al.]%
       {Abhishek K. Srivastava$^1$\thanks{email: \texttt{aks@aries.res.in}},
       V.~M.~Nakariakov$^{2,5}$, B.~N.~Dwivedi$^3$ and Pankaj~Kumar$^{4}$\\
       $^1$Aryabhatta Research Institute of Observational Sciences (ARIES), Manora Peak, Nainital-263 129, India.\\
       $^2$Centre for Fusion, Space and Astrophysics, University of Warwick, Coventry, CV4 7AL, UK.\\
       $^3$Department of Applied Physics, Institute of Technology, Banaras Hindu University, Varanasi-221005, India.\\
       $^4$Korea Astronomy and Space Science Institute (KASI),Hwaamdong, Yuseong-Gu, Daejeon-305-348,
Republic of Korea.\\
       $^5$Central Astronomical Observatory at Pulkovo of the Russian Academy of Sciences, 196140, St Petersburg, Russia.}

\pubyear{2011}
\volume{00}
\pagerange{\pageref{firstpage}--\pageref{lastpage}}

\date{Received \today}

\maketitle
\label{firstpage}

\begin{abstract}
Solar transients and eruptive phenomena which are ubiquitous in the solar
atmosphere, can shed new light to the understanding of the outstanding
problems like coronal
heating and the solar wind acceleration.
Observations in the entire electromagnetic spectrum of such dynamical
processes of large and small-scale transient/eruptive events, with highly dynamic magnetic field configuration, and energetic particles, provide crucial information about the plasma processes at mega-Kelvin temperature embedded in a complex magnetic field, and also energy build-up/energy-release processes, taking place in such events. One of the most important phenomenological aspects of solar eruptive phenomena is the induced magnetohydrodynamic (MHD) waves generated during these energetic processes, which carry a potential
signature to probing the solar active regions. In this paper, we briefly review the recent trends of the transient (e.g., flares) induced quasi-periodic oscillations in the solar atmosphere and
discuss their implications in diagnosing the solar active regions, providing the clue to understanding local plasma dynamics and heating. 
%
%
\end{abstract}


\section{Introduction}\label{s:intro}
Observations of the quasi-periodic oscillations (QPOs) in the solar and stellar atmospheres
are very important phenomena, and have potential applications in diagnosing the 
local plasma conditions of the active/flaring regions to understand their dynamics 
and heating. Recently, the QPOs have been observed associated 
with the solar and stellar flares (Tsap et al., 2011; Nakariakov et al., 2010; Sych et al., 2009; Nakariakov, 2007; Stepanov et al.,2005;
Mitra-Kraev et al., 2005, and references cited there).

\begin{figure}
\centerline{\includegraphics[width=6cm]{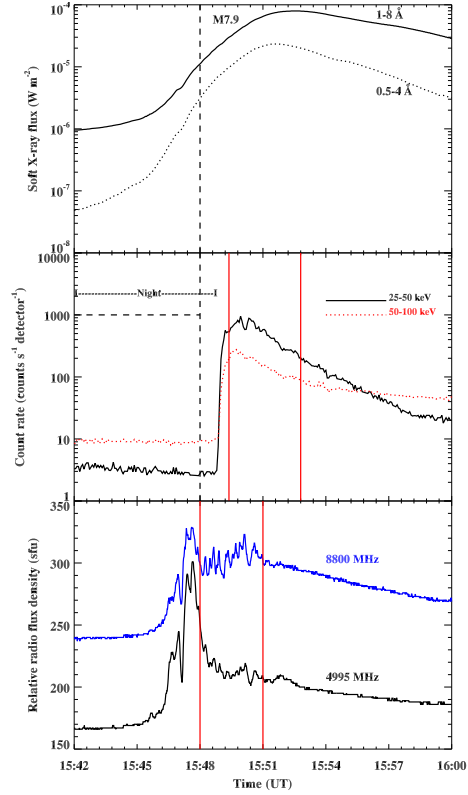}}
\caption{An example of flare associated QPOs : GOES soft X-ray flux, RHESSI hard X-ray and radio flux
profiles for the M7.9/1N flare event on 27 April, 2006. RHESSI
night time period is indicated by dotted line in the second panel.
The decay phase is associated with quasi-periodic oscillations both in HXR and radio fluxes
(adapted from Kumar et al., 2010).
}
\end{figure}

Mostly the quasi-periodic
oscillations in the solar and stellar atmospheres are associated with the transients activities. The strong energy release
in the solar and stellar flares
stored in sheared and twisted magnetic fields are the main
drivers of such observed QPOs in the emissions at various wavelength bands of the emitted electromagnetic spectrum.
It is discovered that 
QPOs associated with the solar
flares can be either generated by various magnetohydrodynamic waves (e.g., Nakariakov and Melnikov, 2009,
and references cited there) 
, or due to the 
periodic changes of the magnetic field configuration and thus periodic 
acceleration of the charged particles 
generated by recurrent magnetic reconnection in the flaring regions (e.g., Jakimiec and Tomczak, 2010 and references cited there)
, or by oscillatory regimes of magnetic reconnection that can be considered as MHD auto-oscillations (Nakariakov et al., 2010). 
The observations of quasi-periodic MHD oscillations 
may be a unique tool to constrain and diagnose the flaring/active region to understand 
its dynamics. However, the 
morphology of the associated loop systems in the active region
must be well known. 

\begin{figure}
\mbox{
\includegraphics[width=8cm]{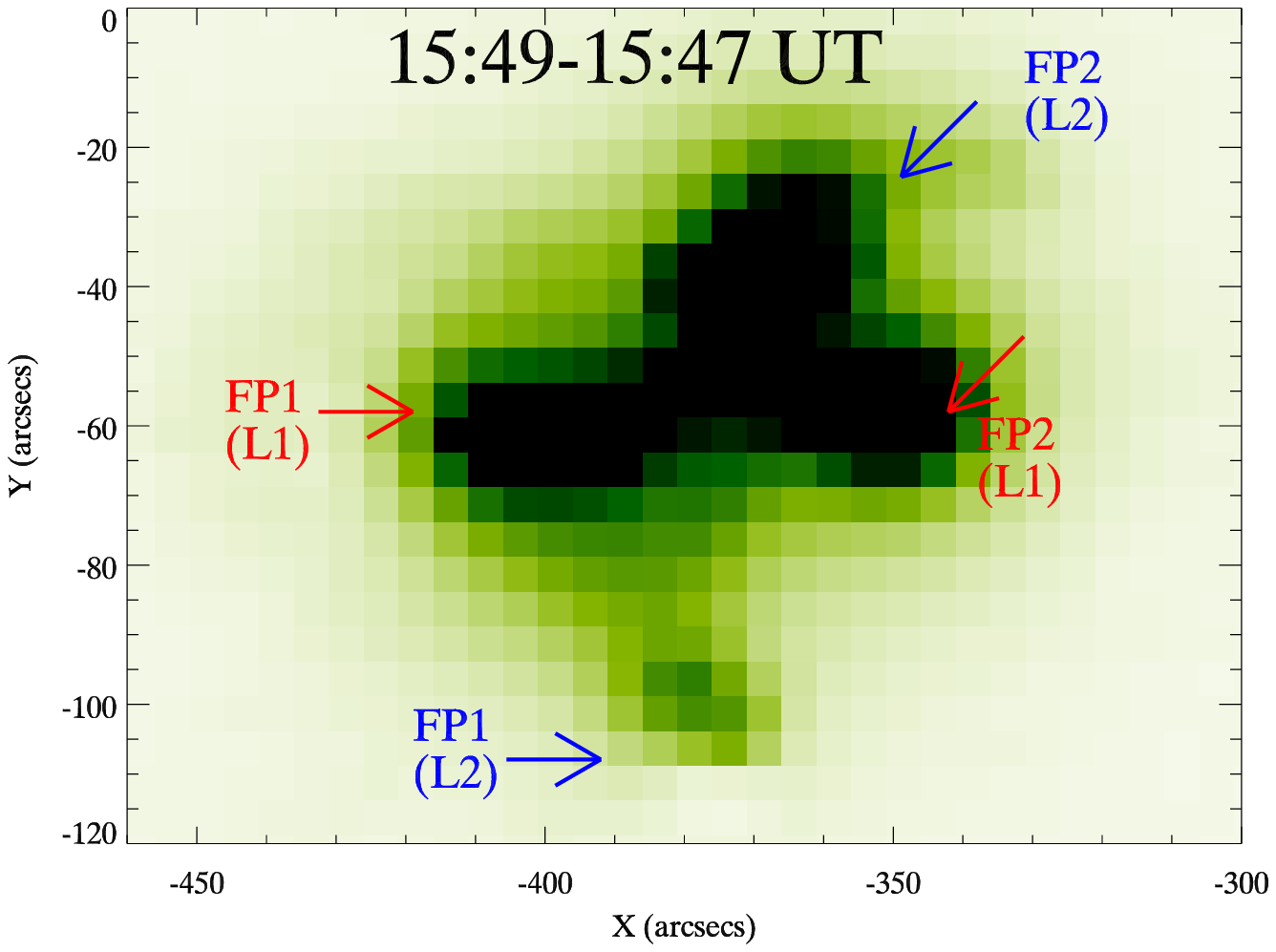}
\includegraphics[width=6cm]{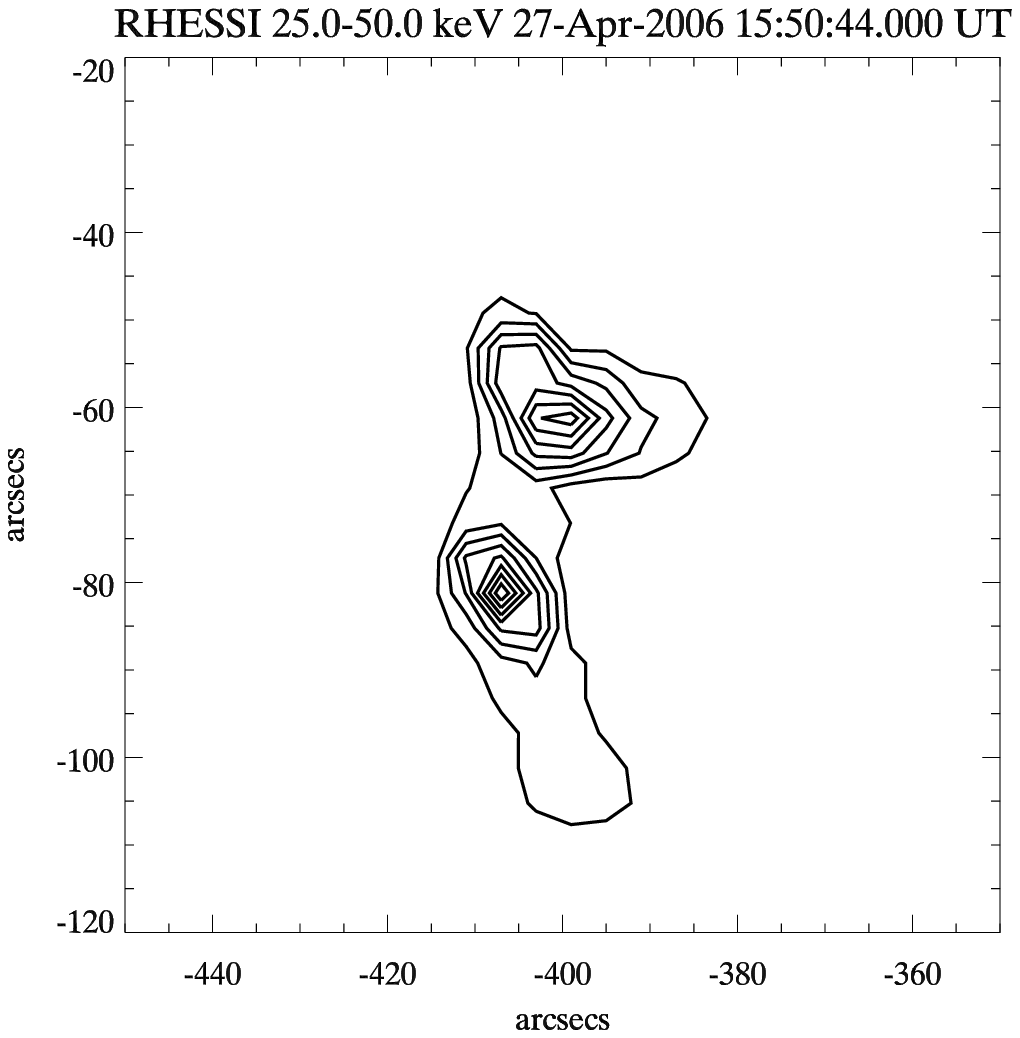}
}
\caption{Left Panel : GOES SXI snapshot on 15:49 UT on 27 April 2006, shows 
the loop-loop interaction during an M7.9/1N flare event. Right Panel :RHESSI
hard X-ray loop-top sources and their merging during their interaction 
is clearly evident on the flare time (adapted from Kumar et al., 2010). This loop-loop interaction is found
to be the most likely reason for the generation of the observed QPOs.
}
\end{figure}

In this brief review paper, we briefly underline some important results related to the
observed QPOs associated with the large-scale solar transients (e.g., flares) in the 
magnetically complex solar active regions, and discuss 
their potential role to diagnose the local plasma conditions.
\section{Some Examples of the Quasi-Periodic Oscillations (QPOs) in the Solar Active Regions}\label{}
A number of observed QPO events in the solar atmosphere are summarized in the on-line archive (cf., http://www.warwick.ac.uk/go/cfsa/people/valery/research/qpp/) that provides a list of
convincingly detected solar QPOs. 
Various significant case studies have been made related to the observations of the QPOs in the solar atmosphere. Foullon et al. (2006) firstly observed the long-period quasi-periodic oscillations of 8-12 min associated with 
the kink mode oscillations of the flaring loops. Nakariakov et al. (2006) have proposed a new model for the
observed long period QPOs of 2-4 min that the fast magnetoacoustic oscillations of a non-flaring loop can interact with a nearby flaring active region, and the part of the oscillation situated outside the loop reaches the regions of steep gradients in magnetic field within an active region to produce periodic variations of electric current density. The variations of the current can induce current-driven plasma micro-instabilities and thus anomalous resistivity that can periodically trigger magnetic reconnection, and hence acceleration of charged particles, producing quasi-periodic pulsations of X-ray, optical and radio emission at the arcade footpoints.
Likewise, reconnection can be triggered by slow magnetoacoustic waves (Chen and Priest, 2006).
A phenomenological relationship between oscillations in a sunspot and quasi-periodic oscillations (QPOs) in flaring energy releases of an active region above the sunspot has been established for the first time by Sych et al. (2009). They have found that the QPOs in the flaring energy releases can be triggered by 3-min slow magnetoacoustic waves leaking from the active region sunspots.
In the flaring event of 2002 July 3 observed by Nobeyama Radioheliograph, Nobeyama Radiopolarimeters, and the RHESSI satellite, Inglis et al. (2009) have discovered the multiple oscillations (28-12 s) associated with the observed QPOs in the spatially resolved emitting region. Nakariakov et al. (2003) have also observed the quasi-periodic loop oscillations of 14-17 s as an important evidence of the global sausage mode oscillations generated by the flare events. In addition to the various observations of the QPOs in hard/soft X-rays, gamma rays, and radio wavelengths (e.g., Nakariakov et al., 2010; Rao et al., 2010; Melnikov et al., 2005; Kleim et al., 2000), these types of
oscillations have also been observed in the white-light as a natural response of the transient phenomena in the lower solar atmosphere (e.g., McAteer et al., 2005). The above mentioned examples and other observed QPO events broadly reveal their origin 
through various physical processes (e.g., MHD oscillation modes, interaction of the large-scale dynamical processes with the compact flaring regions, periodic magnetic reconnections, and the newly proposed Alfv\'enic vortex shedding effects (Gruszecki et al., 2010), etc.) occurred in the solar active/flaring regions at the Sun. The range of these physical processes, therefore, provide 
a unique tool to understand broadly the local physical conditions of the solar atmosphere and solar eruptive regions. 
More accurate knowledge of the source region of these QPOs must be known precisely to understand its connection with the various physical processes.

Recently, Kumar et al. (2010) have reported a first detailed multiwavelength observations of an M-class flare triggered by loop-loop interaction in AR 10875 on 27th April
2006. They have found that a loop coalescence instability may cause the interaction of two multi-temperature loop
systems, which triggers the solar flare.
The M7.9/1N flare occurred on 27 April 2006 in NOAA AR 10875
during 15:45 UT-15:58 UT due to loop-loop interaction (Kumar et al., 2010).
GOES soft X-ray fluxes, RHESSI hard X-ray and radio flux  profiles
for this flare event are shown in Fig. 1. The evidence of
QPOs in 4995,
8800 MHz radio fluxes during $\sim$15:48-15:51 UT
for the duration of $\approx$3 minutes is most likely. The QPOs are also evident  
in the RHESSI HXR emissions of 25-50 and 50-100 keV
energy bands during $\sim$15:49:12-15:52:48 UT. Before 15:49 UT, RHESSI was encountered 
by its night-time phase. However, these QPOs 
are well observed and evident in both HXR and radio frequencies 
in the post flare phase of their heightened emissions.
In the present discussion, we do not aim to analyze in detail the observed quasi-periodic phenomena, and
the main results will be analyzed and published elsewhere. However, the most likely 
association of the QPOs in 
HXR and radio bands, and also its association with the clear morphology of the flaring region
where the loop-loop interaction takes place, may shed new light on the most
possible generation mechanism of such QPOs and its diagnostic capability (cf., Fig. 2). On the first
instance, the origin of QPOs in the present case may be associated with the 
coalescence instability and loop-loop interaction (Tajima et al., 1982; Smartt et al., 1993; Kumar et al., 2010).
However, the possibility of the other mechanisms cannot be ruled out. The observed QPOs
may be due to the periodic change in the length
of the interacting segments of two interacting loops that can result in the periodic
change of the efficiency of the particle acceleration. In such a case, the observed QPOs
will be essentially the non-thermal emissions. However, there is not much 
evidence of it in the present observational base-line.
The observed QPO can also be caused by some periodic regime of magnetic reconnection,
e.g., periodic shedding of the plasmoids (Gruszecki et al., 2010). The reconnection
generated magnetohydrodynamic modes can also be responsible for the generation 
of such oscillations during the flare event (see Nakariakov and Melnikov, 2009
and references cited there).

\section{Discussions and Conclusions}\label{}
In conclusion, the observed parameters of the quasi-periodic oscillations (e.g. the periods, amplitude, spatial and
morphological  informations) must be associated with the physical parameters of
the flaring plasma in the solar active regions and hence can be used for the 
diagnostic purposes through the method
of coronal seismology (see Zaitsev and Stepanov, 2008; Nakariakov and Melnikov, 2009 and references cited there). The QPO
seismology of the solar active regions has various
advantages. The high level of the emissions at various wavelenghts of the electromagnetic spectrum
during the solar eruptive events allows us to
increase the time resolution of the observational instruments. The flare excites various
kinds of magnetohydrodynamic (MHD) waves and oscillations in the compressible, fully ionized, structured and
magnetized plasma, which can be detected in the modulation of the flaring
emissions. Therefore, such MHD modes can be useful for the study of crucial plasma 
parameters of the flaring regions using the principle of MHD seismology 
(e.g., Andries et al., 2009; Aschwanden, 2009; Nakariakov and Melnikov, 2009 and references cited there).
Moreover, periodic and quasi-periodic oscillations are also observed in stellar flares, 
and this opens up interesting
perspectives for stellar coronal seismology and various comparative studies (e.g., Pandey and Srivastava, 2009; 
Mathioudakis et al., 2006, Stepanov et al., 2005).
However, the solar QPOs must be observed with caution and using the multi-instrument and multi-wavelength
emissions because sometime these QPOs may be detected due to the instrumental response (see, Inglis et al., 2011).

The modern space-based (e.g., SDO, Hinode) and ground-based (SST, Nobeyama)
observations are now providing the high spatio-temporal resolution observations of the 
solar atmosphere. The morphology of the solar active regions and associated 
eruptive phenomena are now comparatively well understood, as well as the instrumental
time resolution is now comparatively enough also to catch the short period changes even
in the associated emissions during flare processes at multiwavelength. The  theory of MHD seismology is also in the 
refined stage where by imposing the recent multiwavelength observations 
of the QPOs over it, we can infer the crucial plasma conditions of the flaring region 
e.g., scale heights, equilibrium conditions, density contrast, magnetic field 
divergence factor, magnetic field etc to understand the local plasma dynamics and heating conditions. 
Therefore, the observations of the transient induced quasi-periodic oscillations 
are likely to play an important role in understanding the complexity of
the solar active regions and its plasma dynamics in the current solar cycle and its anticipated intense maximum in
2012-2013.



\appendix
\end{document}